\documentclass[%
 reprint,
 amsmath,amssymb,
 aps,prl,
]{revtex4-2}

\usepackage{graphicx}
\usepackage{dcolumn}
\usepackage{bm}
\usepackage{xcolor}
\usepackage{comment}
\usepackage{multirow}

\defcitealias{Misner:1973prb}{MTW73}

\begin{document}

\title{On the single versus the repetitive Penrose process in a Kerr black hole}

\author{Remo Ruffini$^{1,2,3,4}$, Mikalai Prakapenia$^{1}$, Hernando Quevedo$^{5,1,6}$ and Shurui Zhang$^{1,7,8}$}
 \email{ruffini@icra.it}

 \affiliation{
 $^{1}$ ICRANet, Piazza della Repubblica 10, I-65122 Pescara, Italy }
\affiliation{
$^{2}$ICRANet, 1 Avenue Ratti, 06000 Nice, France
}
\affiliation{
$^{3}$ICRA, Dipartimento di Fisica, Sapienza Universit$\grave{a}$ di Roma, Piazzale Aldo Moro 5, I-00185 Roma, Italy
}
\affiliation{ $^{4}$INAF, Viale del Parco Mellini 84, 00136 Rome, Italy  }

\affiliation{$^{5}$Instituto de Ciencias Nucleares, Universidad Nacional Aut\'{o}noma de M\'{e}xico, AP 70543, Mexico City 04510, Mexico}

\affiliation{$^{6}$Al-Farabi Kazakh National University, Al-Farabi Ave. 71, 050040 Almaty, Kazakhstan}

\affiliation{ $^{7}$School of Astronomy and Space Science, University of Science and Technology of China, Hefei 230026, China}

\affiliation{$^{8}$CAS Key Laboratory for Research in Galaxies and Cosmology, Department of Astronomy, University of Science and Technology of China, Hefei 230026, China}

\begin{abstract}
Extracting the rotational energy from a Kerr black hole (BH) is one of the crucial topics in relativistic astrophysics. Here, we give special attention to the Penrose ballistic process based on the fission of a massive particle $\mu_0$ into two particles $\mu_1$ and $\mu_2$, occurring in the ergosphere of a Kerr BH. Bardeen et al. indicated that for the process to occur, some additional ``{\it hydrodynamical forces or superstrong radiation reactions}'' were needed. Wald and Chandrasekhar further expanded this idea. This animosity convinced T. Piran and collaborators to move from a simple three-body system characterizing the original Penrose process to a many-body system.  This many-body approach was further largely expanded by others, some questionable in their validity. Here, we return to the simplest original Penrose process and show that the solution of the equations of motion, imposing the turning point condition on their trajectories, leads to the rotational energy extraction from the BH expected by Penrose. The efficiency of energy extraction by a single process is quantified for three different single decay processes occurring respectively at $r=1.2 M$,  $r=1.5 M$, and $r=1.9 M$. An interesting repetitive model has been proposed by Misner, Thorne \& Wheeler \citep[][hereafter MTW73]{Misner:1973prb}. Indeed, it would appear that a repetitive sequence of $246$ decays of the above injection process at $r=1.2 M$ and the corresponding ones at $r=1.5 M$ and $r=1.9 M$ could extract $100\%$ of the rotational energy of the BH, so violating energy conservation. The accompanying paper, accounting for the existence of the BH irreducible mass, introduces a non-linear approach that avoids violating energy conservation and leads to a new energy extraction process.
\end{abstract}

\date{\today}

\maketitle

\section{\label{sec:level1} 1. Introduction}

The physics of a Kerr Black Hole (BH) is based on three contributions:
\begin{enumerate}
\item The mathematical solution of the Einstein equations by Roy P. Kerr \cite{1963PhRvL..11..237K} which plays a central role in describing rotating objects in General Relativity.
\item The discovery by Carter of the separability of the Hamilton-Jacobi equations, describing geodesics in the Kerr spacetime, in terms of four conserved quantities \cite{1968PhRv..174.1559C,landau2013classical}.
\item The introduction by Ruffini and Wheeler of the effective potential governing the motion of test particles in the BH equatorial plane \cite{landau2013classical,1971ESRSP..52...45R}
\begin{gather}
\frac{V}{\mu}
= \frac{2 \hat{a} \hat{p}_{\phi} \pm \sqrt{\hat{r}\left[\hat{a}^2+\hat{r}\left(\hat{r}-2\right)\right][\hat{p}_{\phi}^2\hat{r}+\hat{r}^3+\hat{a}^2(\hat{r}+2)]}}{\hat{r}^3+\hat{a}^2\left(\hat{r}+2\right)}.
\label{eq:effpMass}
\end{gather}
\end{enumerate}
We use geometrical units throughout and consider the positive sign solution of Eq. (\ref{eq:effpMass}). 
For the BH, we indicate by $M$ and $L$ the mass and angular momentum and by $\hat a = L/M^2$ the dimensionless spin parameter. Similarly, for a massive test particle of mass $\mu$ and angular momentum $p_\phi$ and energy $E$, we indicate the dimensionless quantities $\hat{E} = E/\mu$, $\hat{p}_{\phi} = p_{\phi}/(M \mu)$, and for the dimensionless radial coordinate is given by $\hat{r} = r/M$. 

A particle  with energy $E$ and $E>V$ is endowed with a positive kinetic energy with 
 $\left( \frac{dr}{ds} \right)^2 > 0$. The particles with $E=V$ correspond to the turning point with $\frac{dr}{ds}=0$.
For stable circular orbits, the conditions $V' = 0$ and $V''>0$ are satisfied, where a prime denotes derivative relative to the radial coordinate (see \citet{landau2013classical,1971ESRSP..52...45R}). 

From the above equations, it was possible to define the new concept of the {\it ergosphere} 
(see, e.g., \citet{landau2013classical,1971ESRSP..52...45R,1971PhT....24a..30R}), which extends from the horizon $ \hat r_+ = 1 + \sqrt{1-\hat a ^2}$ to the stationary limit $\hat r _{\rm ergo} = 1 + \sqrt{1-\hat a ^2 \cos^2\theta}$.

In 1969, Penrose proposed \cite{1969NCimR...1..252P} the occurrence of a decay process between the horizon and the stationary limit of a Kerr black hole (BH) to extract its rotational energy. In this approach, a massive particle $\mu_0$ with energy $E_0$ decays into two particles $\mu_1$  and $\mu_2$. The particle $\mu_1$, with negative energy $E_1$ and negative angular momentum $p_{\phi 1}$, enters into the BH horizon, while the particle $\mu_2$ reaches back a distance at infinity with total energy $E_2$ larger than $E_0$. The difference of energy between $E_2$ and $E_0$ corresponds to the extraction of the rotational energy of the BH.
 
The first process example was in Fig.~2 of Christodoulou's paper 
\cite{1970PhRvL..25.1596C} (submitted on September 17, 1970), which summarizes the computation by Ruffini and Wheeler in \cite{1971ESRSP..52...45R},
where some kinematic parameters were determined. This result was possible by assuming that the decay occurs at the three turning points of the particles' trajectories. The equations from which this example was derived and the conditions of the turning point were later published  \cite{1971ESRSP..52...45R}. Three months after the publication of Christodoulou's paper, the Penrose-Floyd paper appeared \cite{1971NPhS..229..177P} (submitted on December 16, 1970). The same equations were solved, missing the crucial simplification of the turning points. Therefore, only a general statement was inferred ``{\it The particular example described here is somewhat inefficient. }''.

Following this line, many papers appeared, the most quoted by Bardeen et al. \cite{bardeen1972rotating}. Instead of solving the equations of the potential as seen from infinity, they went through a detailed analysis of the description of the trajectories by using an orthonormal tetrad formalism around every spacetime event. However, this work missed verifying the actual trajectories. 
They concluded that for the Penrose process to be viable, the outgoing particle should have a velocity $> c/2$. Bardeen et al. stated  
that ``... {\it energy extraction cannot be
achieved unless hydrodynamical forces or superstrong radiation reactions can accelerate fragments to more than this speed during the infall. On dimensional grounds,
such boosts seem to be excluded} ...''.
Chandrasekhar made an analogous statement in \cite{chandrasekhar1985mathematical}, page 371: ``{\it In other words, the fragments must have relativistic energies before any extraction of energy by a Penrose process becomes possible}''.  
 
This led Piran et al. \cite{piran1975high,leiderschneider2016maximal,kovetz1975efficiency} to consider the many-particle Penrose collisional process. This multi-particle approach was correctly performed in \cite{piran1975high,leiderschneider2016maximal,kovetz1975efficiency}. 

Indeed, as shown below, the simplest Penrose process can fulfill the above inequality, which does not constrain its validity{, and} generalized processes are not necessary to prove its feasibility: {regular fission works fine}.

\begin{figure*}
	\includegraphics[width=\hsize]{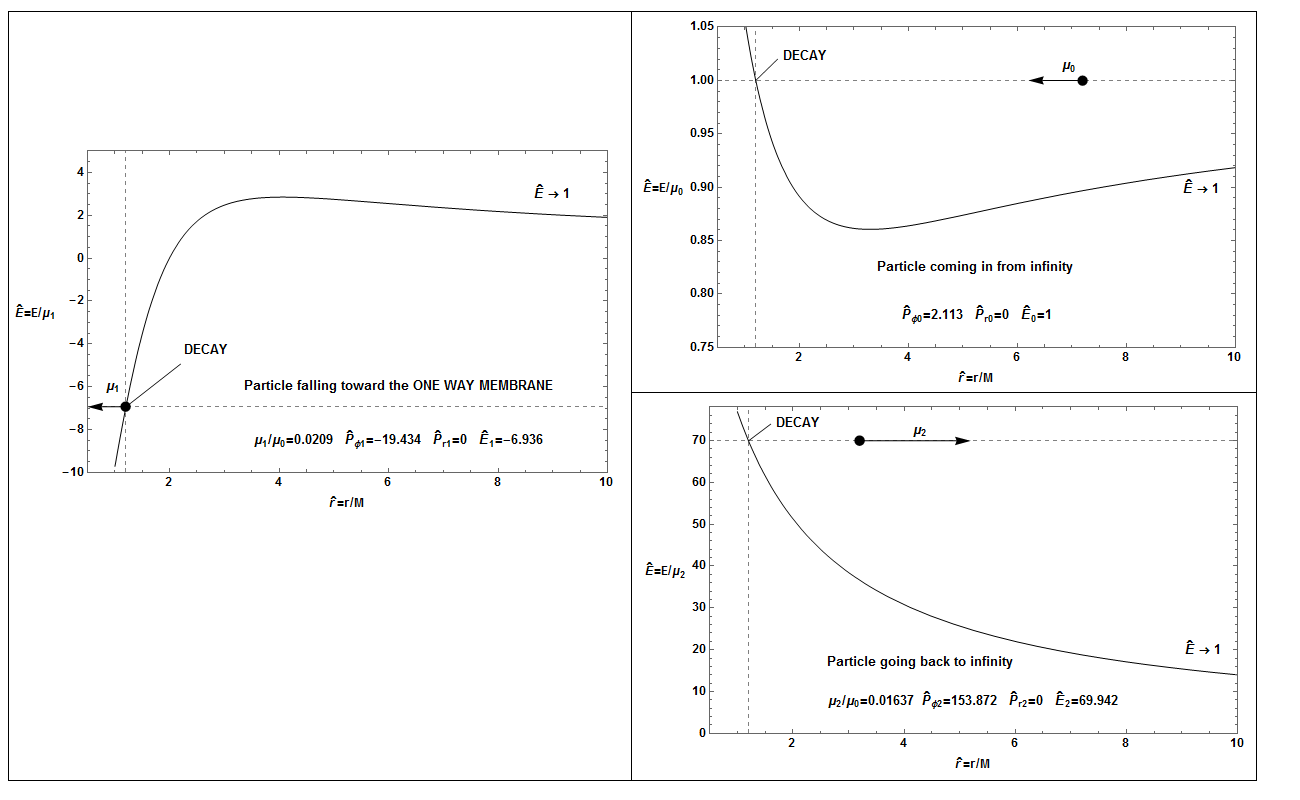}
	\caption{Example of a particle decay into two particles in the {\it ergosphere} of an extreme Kerr BH occurring at $\hat{r}=1.2$. The three particles fulfill the mass condition for $\mu_1/\mu_0=0.0209$ and $\mu_2/\mu_0=0.01637$. For these masses, and further assuming $\hat{E}_0=1.0$ and $\hat{p}_{\phi1}=-19.434$, the solution of the conservation equations given in Sec.~\ref{sec:level2} leads to $\hat{p}_{\phi0}=2.113$, $\hat{E}_1=-6.936$,  $\hat{E}_2=69.942$, and $\hat{p}_{\phi2}=153.872$. The \textbf{upper right panel} shows the particle ``0'' coming in from infinity. The \textbf{left panel} displays particle ``1'' with negative energy (represented by a black point) falling towards the horizon. The \textbf{lower right panel} shows the particle ``2'' returning to infinity after decay.}
	\label{fig:threeTP12}
\end{figure*}

The paper is organized as follows. For general convenience, we first formulate the energy and angular momentum conservation laws, assuming the essential condition that the decay occurs at the turning points of the effective potential \cite{1970PhRvL..25.1596C}. {The case of an extreme Kerr BH with $a/M=1$ is assumed}. Then, we obtain the solution of the equations for an incoming particle $\mu_0$ with mass $10^{-2}M$ that decays close to the horizon at $\hat r =1.2$. These choices respond to two opposite requirements: to test the most energetic process by performing the decay close to the horizon and fulfill the test particle approximation for the incoming particle, see Misner, Thorne \& Wheeler \citep[][hereafter MTW73]{Misner:1973prb}. We determine the masses $\mu_1$ and $\mu_2$ and other relevant quantities the conservation laws imply. {The solution for any other value of $\mu_0$ can be obtained by simply rescaling the quantities.} We focus on the change of the BH mass and angular momentum due to the capture of particle $\mu_1$ and the relativistic nature of the particle $\mu_2$, showing that it is ultra-relativistic and carries away a portion of the BH energy and angular momentum, fulfilling the {Penrose expectation. We verify} the decay process in three locations: $1.2M$, $1.5M$, and $1.9M$. The results are summarized in Fig.~\ref{fig:threeTP12} and Table~\ref{tab:table1}. 

Subsequently, following a suggestion by \citepalias{Misner:1973prb}, we perform an iterative Penrose process and verify {if} in all three above cases, it is possible to extract the entire rotational energy of a Kerr BH in a finite number of iterative steps (see Fig.~\ref{figlinear12}). However, to perform this process, the injected particles have a much larger mass than the BH, which {violates the conservation of the} mass-energy {of the system}. We are then confronted with a basic difficulty in implementing {correctly the} repetitive Penrose process. This problem is addressed in a {companion} publication \cite{2024arXiv240510459R} which implements the BH mass-energy formula \citep{1970PhRvL..25.1596C,1971PhRvD...4.3552C,1971PhRvL..26.1344H}, accounting for the irreducible mass $M_{\rm irr}$ and its relation to the horizon surface area $S=16\pi M_{\rm irr}^2$. This will solve the above inconsistency, leading to a highly nonlinear energy extraction process.

\section{\label{sec:level2} 2. The Penrose process basic equations}

We denote the intervening particle masses by $\mu_0$, $\mu_1$, and $\mu_2$, with {the process mass defect fulfilling
\begin{equation}
\mu_{\rm defect} \equiv \mu_0-\mu_1-\mu_2 > 0,
\end{equation}
implied by nonlinear general relativistic processes}.

Following \citepalias{Misner:1973prb}, we assume that the test-particle approximation holds, so we use the mass ratio $\mu_0/M\leq 10^{-2}$. We assume that the decay process occurs at the turning points in the {\it ergosphere} of an extreme Kerr BH, $a/M=1$. The particle ``1'' counter-rotates with the BH and is in a negative energy state, as measured from an infinite distance \cite{landau2013classical,1971ESRSP..52...45R,1971PhT....24a..30R,1970PhRvL..25.1596C,bardeen1972rotating,chandrasekhar1985mathematical}. Particle ``2'' returns to infinity with mass-energy $\mu_2 \hat{E}_2>\mu_0 \hat{E}_0$, resulting in the extraction of rotational energy from the BH. 

The effective potential is deduced from Eq. (\ref{eq:effpMass}) for $\hat a = 1$, which for $r/M\to 1$, $V/\mu\to \hat{p}_{\phi}/2$. For particle motion at $\hat r > \hat r_+ = 1$, it leads to $\hat E >\hat p_{\phi}/2$. By setting values of $\hat{r}$, $\hat{E}_1$ and $\hat{p}_{\phi1}$ that fulfill $1<\hat{r}<2$, $\hat{E}_1<0$, $\hat{p}_{\phi1}<0$, $\hat{E}_1>\hat{p}_{\phi1}/2$, and masses that satisfy the above-mentioned mass condition, one can find the corresponding positive values of geodesic conserved quantities $\hat{E}_0$, $\hat{p}_{\phi0}$, $\hat{E}_2$, and $\hat{p}_{\phi2}$ from the set of equations  \citep{1974bhgw.book.....R}
\begin{subequations}\label{eq:eq6}
    \begin{align}
\mu_0\hat{E}_0&=\mu_1\hat{E}_1+\mu_2\hat{E}_2,\label{subeq:eq6b}\\
\mu_0\hat{p}_{\phi0}&=\mu_1\hat{p}_{\phi1}+\mu_2\hat{p}_{\phi2},\label{subeq:eq6c}\\
    \mu_0 \hat{p}_{r0} &= \mu_1 \hat{p}_{r1} + \mu_2\hat{p}_{r2},
    \label{subeq:eq6d}
    \end{align}
\end{subequations}
and
\begin{equation}\label{eq:pmus}
    (\hat{E}_i^2-1)\hat{r}^3+2\hat{r}^2 +(\hat{E}_i^2-\hat{p}_{\phi i}^2-1)\hat{r}+2(\hat{E}_i-\hat{p}_{\phi i})^2=0,
\end{equation}
where Eqs. (\ref{subeq:eq6b})--(\ref{subeq:eq6d}) are the conservation of energy, angular momentum, and radial momentum, respectively. Equations (\ref{eq:pmus}) guarantee the particles {$i=0$, $1$, and $2$} are at a turning point. The above system of algebraic equations (\ref{subeq:eq6b})-(\ref{eq:pmus}) is {undetermined}. However, we are not free to select any values for five unknowns arbitrarily since they must satisfy the process boundary conditions. We address only the maximum energy extraction process in agreement with the test-particle approximation.

\begin{table*}
\caption{\label{tab:table1}%
Changes of the parameters of an initially extreme Kerr BH with mass $M$ and angular momentum $L=M^2$ by capturing the particle with negative energy $E_1=\hat E_1 \mu_1=\Delta M$ and negative angular momentum $\Delta L=\hat p_{\phi 1}\mu_1 M $. We list the results for the decay at three positions inside the ergosphere: $\hat{r}=1.2$, $1.5$, and $1.9$. We choose {$\hat E_0 = 1$, $\hat p_{\phi1} =-19.434$, $\mu_2/\mu_1 = 0.78345$, and from the conservation laws, we obtain the other parameters. For $\hat r=1.2$: $\mu_1 = 0.0209\mu_0$, $\mu_2 =0.01637\mu_0$, $\hat p_{\phi0} = 2.113$, $\hat E_1 = -6.936$, $\hat E_2 = 69.942$, and $\hat p_{\phi2} = 153.872$. For $\hat r=1.5$: $\mu_1 = 0.0218\mu_0$, $\mu_2 =0.0171\mu_0$, $\hat p_{\phi0} = 2.268$, $\hat E_1 = - 3.520$, $\hat E_2 = 62.995$, and $\hat p_{\phi2} = 157.485$. For $\hat r = 1.9$: $\mu_1 = 0.0247\mu_0$, $\mu_2 =0.0194\mu_0$, $\hat p_{\phi0} = 2.456$, $\hat E_1 = -0.501$, $\hat E_2 = 52.326$, and $\hat p_{\phi2} = 151.737$. For all cases, we adopt $\mu_0 = 10^{-2}M$.}}
\begin{ruledtabular}
\begin{tabular}{c|ccccccc}
$\hat r$ & $\displaystyle\frac{E_2}{M}$ & $\gamma_2$ & $\displaystyle\frac{E_{\rm extracted}}{M}$ & $\displaystyle\frac{E_1}{M} = \displaystyle\frac{\Delta M}{M}$ & $\displaystyle\frac{\Delta L}{M^2}$ & $\displaystyle\frac{L_{f}}{M^2}$ & $\displaystyle\frac{M_{f}}{M}$ \\[6pt]
\hline 
$1.2$ & $1.145 \times 10^{-2} $   & $ 69.942$ & $ 1.449 \times 10^{-3}$ & $-1.449 \times 10^{-3} $ & $ -4.061 \times 10^{-3}$ & $ 0.9959 $ & $ 0.9985 $ \\
$1.5$  & $1.076 \times 10^{-2} $   & $ 62.99$ & $ 7.681 \times 10^{-4}$ & $-7.681 \times 10^{-4} $ & $ -4.240 \times 10^{-3}$ & $ 0.9957 $ & $ 0.9992 $ \\
$1.9$ & $1.012 \times 10^{-2} $  & $ 52.326$ & $ 1.236 \times 10^{-4}$ & $-1.236 \times 10^{-4} $ & $ -4.799 \times 10^{-3}$ & $ 0.9952 $ & $ 0.9998 $ 
\end{tabular}
\end{ruledtabular}
\end{table*}

When the particle ``1'' falls into the BH, it reduces the mass and angular momentum of the BH according to
\begin{equation}
    \Delta M = \hat{E}_1\mu_1, \quad \Delta L = \hat{p}_{\phi1}M\mu_1,
    \label{eq:dM}
\end{equation}
and the energy extracted from the BH by particle ``2'' is
 \begin{equation}
    E_{\rm extracted} = \hat{E}_2\mu_2 - \hat{E}_0\mu_0 = - \hat E_1 \mu_1.
    \label{eq:Eextracted}
\end{equation}

\section{\label{sec:level4} 3. Consequences of the energy extraction process}

A new simplifying approach avoids numerical procedures. We adopt the particle ``0'' drops from rest at infinity ($\hat E_0 =1$), set the radius $\hat r$, and supply a value of $\hat p_{\phi1}$ and the outgoing to infalling particle mass-ratio $\mu_2/\mu_1$. In this way, the system of equations can be solved analytically for the rest of variables (see details in \cite{2024arXiv240510459R}):
\begin{subequations}\label{eqs:solution}
    \begin{align}
        \hat p_{\phi0} &= \frac{2-(\hat r-1)\sqrt{2 \hat r}}{2-\hat r},\quad \hat p_{\phi2} = \frac{\mu_0}{\mu_2}\hat p_{\phi0}-\frac{\mu_1}{\mu_2}\hat p_{\phi1},\\
        \hat E_1 &= \frac{2\,\,\hat p_{\phi1} + \sqrt{\rho+\hat r^2 \Delta\,\hat p_{\phi1}^2}}{R},\quad \hat E_2 = \frac{\mu_0}{\mu_2}-\frac{\mu_1}{\mu_2}\hat E_1,\\
        \mu_1 &= \mu_0 \frac{\rho}{K + \sqrt{K^2-\rho^2 [1-(\mu_2/\mu_1)^2]}}
    \end{align}
\end{subequations}
where 
\begin{subequations}\label{eqs:defs}
    \begin{align}
        \Delta &= (\hat r-1)^2,\quad R=\hat r^3 + \hat r+2,\quad \rho = \hat r\,\Delta\,R\\
        K &= \sqrt{(\rho + \hat r^2 \Delta \hat p_{\phi0}^2)(\rho + \hat r^2 \Delta \hat p_{\phi1}^2)}-\hat r^2 \Delta \hat p_{\phi0} \hat p_{\phi1}.
    \end{align}
\end{subequations}

Table~\ref{tab:table1} and Fig.~\ref{fig:threeTP12} summarize the results from Eqs. (\ref{eqs:solution}) and (\ref{eqs:defs}), assuming $\hat p_{\phi1} =-19.434$ and $\mu_2/\mu_1 = 0.78345$, for three selected decay points, $\hat r = 1.2$, $1.5$, and $1.9$. We can use these values to determine the new BH parameters, i.e., $M_f = M_i + E_1 = M + \hat E_1\mu_1 = 0.9985 M$ and $L_f = L_i + L_1= M^2 + \hat p _{\phi 1} \mu_1 M = 0.9959 M^2  $. Finally, the extracted energy becomes $E_{\rm extracted}=E_2 - E_0 = \hat E_2 \mu_2 - \mu_0=1.45\times 10^{-3}M$. 

Notice that the energy of the incoming and outgoing particles can be written as $E_0 = \gamma_0 \mu_0$ and $E_2 = \gamma_2 \mu_2$, so $\hat E_0$ and $\hat E_2$ are the Lorentz factors $\gamma_0$ and $\gamma_2$ {at infinity}. Thus, $\gamma_0=1$, while the particle ``2'' that carries out the rotational energy, $\gamma_2 \simeq 69.942$, which corresponds to a velocity of approximately $99.989\%$ of the speed of light, i.e., the outgoing particle is ultra-relativistic. This means that, in this particular case, the outgoing particle acquires a clear signature with an enormous kinetic energy due to its interaction with the gravitational field of the BH. It is easy to verify that this result satisfies the inequalities established by Wald \cite{wald1974energy}, and Bardeen et al. \cite{bardeen1972rotating} (see also \cite{chandrasekhar1985mathematical}).

For $\hat r = 1.5$, we obtain $\mu_1/\mu_0 = 0.0218$, $\mu_2/\mu_0=0.0171$, $\hat p_{\phi 0} = 2.268$, $\hat E _ 1 = - 3.520$, $\hat p_{\phi 2} = 157.485$ $\hat E _ 2 = 62.995$. For $\hat r = 1.9$, we obtain $\mu_1/\mu_0= 0.0247$, $\mu_2/\mu_0 =0.0194$, $\hat p_{\phi 0} = 2.456$, $\hat E _ 1 = - 0.501$, $\hat E _ 2 = 52.326$, and $\hat p_{\phi 2} = 151.737$. A summary of the main quantities for the above three decays is given in Table \ref{tab:table1}.

The Penrose process remains the only consistent theoretical model to extract the rotational energy of {an extreme Kerr BH with $a/M=1$ using massive test particles} in the BH ergosphere with an efficiency larger than nuclear or chemical energy processes \cite{1927Sci....65..431M}. Indeed, defining the process efficiency by $\eta \equiv E_{\rm extracted}/E_0$, we obtain efficiencies of $14.5\%$, $7.7\%$, and $1.2\%$ for the three cases $\hat r =1.2$, $1.5$, and $1.9$. The corresponding multi-particle approach by Piran reaches an efficiency of $12.6\%$ \cite{piran1975can,kovetz1975efficiency,leiderschneider2016maximal}.  

The main problem with the above procedure is that {the original Penrose process is a linear one: it neglects the nonlinearities introduced by a varying BH spin, implying the transition of the effective potential from Eq.(\ref{eq:pmus}) to Eq. (\ref{eq:effpMass}); it neglects as well the nonlinearities introduced by the mass-energy formula \citep{1971PhRvD...4.3552C,1971PhRvL..26.1344H}. Both these effects are mandatory} as soon as the repetitive processes are considered (see below).

\section{4. A repetitive linear process}
\label{sec:rep}

A generalization of the Penrose process was suggested in Fig.~33.2 of \citepalias{Misner:1973prb}, moving from a single process to a repetitive one, {invoking a scale-invariant process}. The authors {follow} the program imagined by Penrose of ``a civilization which has built some form of stabilized structure surrounding the Black Hole'' \citep[see Fig.~5 in Ref.][]{1969NCimR...1..252P}, by proposing ``{\ An advanced civilization has constructed a rigid framework around a black hole, and has built a huge city on that framework. Each day trucks carry one million tons of garbage out of the city to the garbage dump. At the dump the garbage is shoveled into shuttle vehicles which are then, one after another, dropped toward the center of the black hole}''. In terms of the Penrose process, it implies repeating decays of the same kind, one by one, into the BH.

\begin{figure}
    \centering
    \includegraphics[width=\hsize]{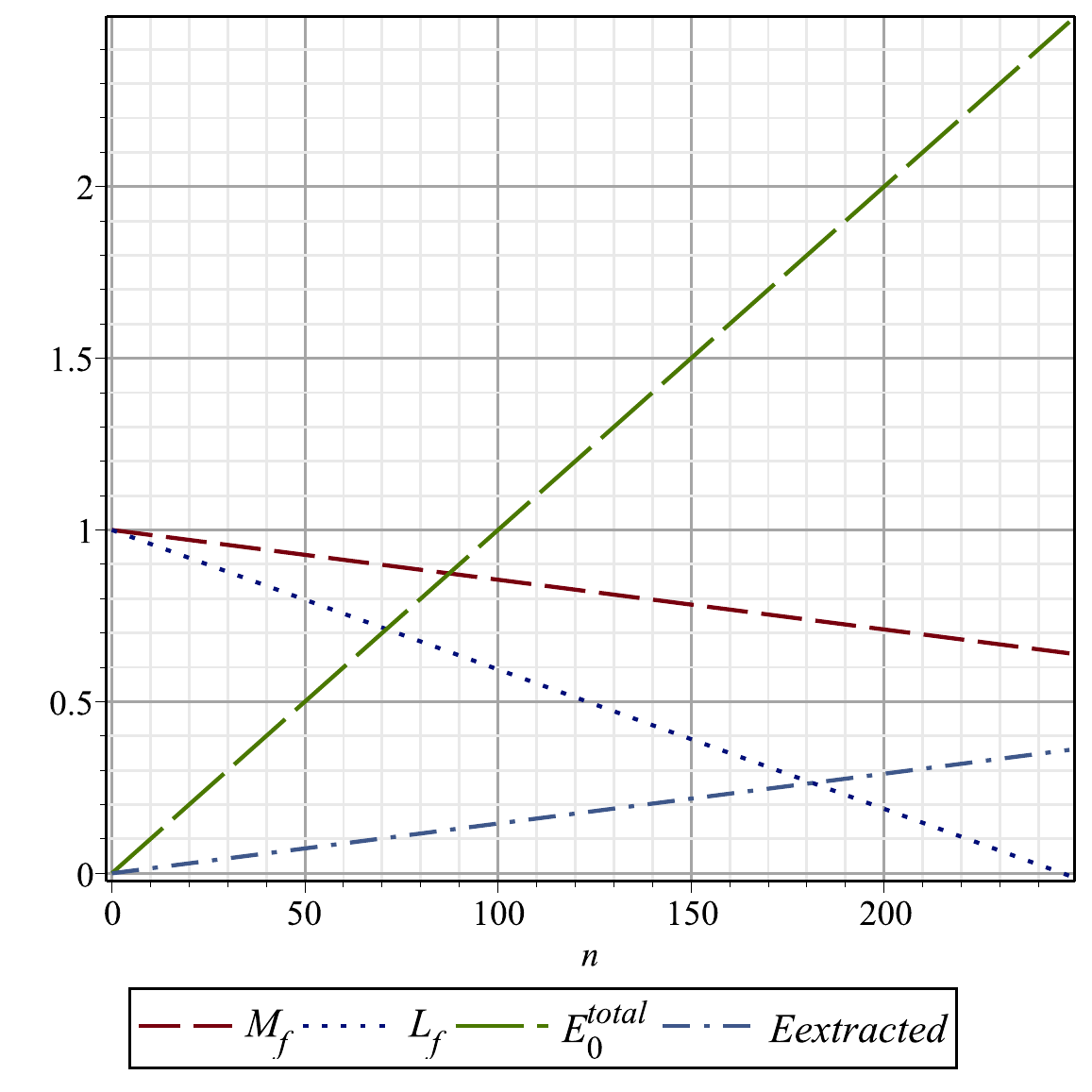}
    \caption{A linear repetitive decay process occurring at $\hat r = 1.2$ A number $n$ of particles with energy $E_0$ is injected into the BH. After $n\approx 246$ decays, the end state is a Schwarzschild BH of mass $\sim 0.643M$.}
    \label{figlinear12}
\end{figure}

The {original} Penrose process is linear in the angular momentum and the mass, i.e., for a given BH initial state $M_i$ and $L_i$, the capture of a particle with energy $E_1$ and angular momentum $p_{\phi 1}$ leads to $M_f = M_i + E_1 $ and $L_f = L_i + p_{\phi 1}$. Therefore, if one were to implement the advanced civilization idea by repeating $n$ times the capture process with the same particle with energy $E_1$ {(see Fig.~\ref{fig:threeTP12})}, one could {reach the BH final state $M_f(n) = M_i + n E_1$ and $L_f(n) = L_i + n p_{\phi 1} = 0$. This clearly violates mass-energy conservation (see Fig.~\ref{figlinear12})}.

These paradoxes prove the incompleteness of the above linear implementation of repetitive Penrose processes. To fulfill the mass-energy conservation, the solution must consider the mass-energy formula (see Eq.~\ref{eq:massformula}), which relates nonlinearly {the mass $M$, the angular momentum $L$}, as well as the irreducible mass $M_{\rm irr}$. Furthermore, {also to be fulfilled is the BH surface area} increase law \citep{1970PhRvL..25.1596C,1971PhRvD...4.3552C,1971PhRvL..26.1344H}, i.e., the BH reversible and irreversible transformations have to be considered. The concept of {\it extractable energy}, complementary to the {\it extracted energy} in the original Penrose process computed above, has to be {implemented}. The decay process becomes dominated by {nonlinearities and any} scale invariance breaks down. 

\section{\label{sec:level5} 5. Conclusions}
\label{sec:conc}

We have revisited the {linear} Penrose process basics and shown that, {in the case $a/M=1$}, it remains a remarkably simple method to probe the Kerr BH rotational energy extraction. We perform this process at three radii within the ergosphere, $\hat{r}=1.2$, $\hat{r}=1.5$, and $\hat{r} = 1.9$. We overcome previous criticisms toward the original Penrose process, confirming it extracts the expected energy and angular momentum with an efficiency of $14.5\%$ in the single process.

We have then considered the proposal by \citepalias{Misner:1973prb} of a repetitive Penrose process. Since the Penrose process is linear in mass and angular momentum, one may consider performing such an iteration assuming that the energy and angular momentum extracted at each step coincide with those of the first step. Under this assumption, {the entire} rotational energy can be extracted (see Fig.~\ref{figlinear12}). However, energy conservation is violated. Indeed, the extracted energy should be computed at each step using the new $a/M$ derived from the new trajectories in the Kerr solution {given by Eq.(\ref{eq:effpMass})}. This introduces a non-linearity in the original Penrose process. This path is approached in the accompanying paper \cite{2024arXiv240510459R}, where it is proved that such nonlinearities are equivalent to introducing the BH mass-energy formula \citep{1971PhRvD...4.3552C,1971PhRvL..26.1344H}
\begin{equation}\label{eq:massformula}
    M^2=M_{\rm irr}^2+\frac{L^2}{4 M_{\rm irr}^2}, \quad S=16 \pi M_{\rm irr}^2
\end{equation}
relating the BH parameters {$M$, $L$, and $M_{\rm irr}$}. The irreducible mass $M_{\rm irr}$ directly relates to the increasing BH surface {area $S$}. In \cite{2024arXiv240510459R}, the irreversibility of the Penrose process is proved, introducing the highly nonlinear {\it extractable energy}, in addition to {the} {\it extracted energy} discussed in the first part of this paper. Crucial will be to compare and contrast the decays at $\hat{r}=1.2$, $\hat{r}=1.5$, $\hat{r}=1.9$ {in such a new approach}. We use the Penrose process as a prototype to infer how the irreducible mass significantly affects {any} rotational energy extraction process in {pure} self-gravitating systems.

\begin{acknowledgments}
{We like to express our gratitude to the two anonymous referees for their outstanding reports and important suggestions.}
We thank Profs. C.L. Bianco,  D. Christodoulou, R. Kerr, and J.A. Rueda for discussions on this work.
{The work of HQ was supported by UNAM PASPA-DGAPA.}
\end{acknowledgments}

\end{document}